\documentclass[12pt]{iopart}

\usepackage{amssymb}
\usepackage{graphicx}
\usepackage[breaklinks=true,colorlinks=true,linkcolor=blue,urlcolor=blue,citecolor=blue]{hyperref}
\usepackage[utf8]{inputenc}

\usepackage[margin=1in]{geometry}

\usepackage{subfigure}
\usepackage{multicol}
\usepackage{multirow}

\usepackage{bm}

\begin{document}

\title[Critical initial-slip scaling for the noisy complex Ginzburg--Landau equation]
{Critical initial-slip scaling for the noisy complex Ginzburg--Landau equation}

\author{Weigang Liu and Uwe C. T\"auber}

\address{Department of Physics \& Center for Soft Matter and Biological Physics, MC 0435,
Robeson Hall, 850 West Campus Drive, Virginia Tech, Blacksburg, VA 24061, USA}

\ead{qfsdy@vt.edu, tauber@vt.edu}


\begin{abstract}
We employ the perturbative field-theoretic renormalization group method to investigate the 
universal critical behavior near the continuous non-equilibrium phase transition in the
complex Ginzburg--Landau equation with additive white noise. 
This stochastic partial differential describes a remarkably wide range of physical systems:
coupled non-linear oscillators subject to external noise near a Hopf bifurcation instability;
spontaneous structure formation in non-equilibrium systems, e.g., in cyclically competing 
populations; and driven-dissipative Bose--Einstein condensation, realized in open systems on 
the interface of quantum optics and many-body physics, such as cold atomic gases and 
exciton-polaritons in pumped semiconductor quantum wells in optical cavities.
Our starting point is a noisy, dissipative Gross--Pitaevski or non-linear Schr\"odinger 
equation, or equivalently purely relaxational kinetics originating from a complex-valued 
Landau--Ginzburg functional, which generalizes the standard equilibrium model A critical
dynamics of a non-conserved complex order parameter field. 
We study the universal critical behavior of this system in the early stages of its relaxation
from a Gaussian-weighted fully randomized initial state.
In this critical aging regime, time translation invariance is broken, and the dynamics is
characterized by the stationary static and dynamic critical exponents, as well as an 
independent `initial-slip' exponent. 
We show that to first order in the dimensional expansion about the upper critical dimension, 
this initial-slip exponent in the complex Ginzburg--Landau equation is identical to its 
equilibrium model A counterpart. 
We furthermore employ the renormalization group flow equations as well as construct a 
suitable complex spherical model extension to argue that this conclusion likely remains true 
to all orders in the perturbation expansion.    
\end{abstract}


\vspace{2pc}
\qquad\qquad\quad 
         driven-dissipative Bose--Einstein condensation, renormalization group

\submitto{\JPA --- \today}

\section{Introduction}

Physical systems display characteristic singularities when their thermodynamic parameters 
approach a critical point. 
The ensuing singular behavior of various observables are governed by a considerable degree 
of universality: critical exponents and amplitude ratios are broadly independent of the 
microscopic details of the respective systems.
The emergence of scale invariance, associated thermodynamic singularities, and universality 
near continuous phase transitions is theoretically understood and described by the
renormalization group (RG), which also allows a systematic computation of critical exponents 
and associated scaling functions (see, e.g., Refs.~\cite{Amit84, Itzykson89, Kleinert01,
Zinn05}).
These concepts and theoretical tools can be extended to dynamical critical behavior near
equilibrium, which may similarly be grouped into various dynamical universality classes. 
In addition to global order parameter symmetries, the absence or presence of conservation 
laws for the order parameter and its coupling to other slow conserved modes crucially 
distinguish dynamical critical properties \cite{Hohenberg77, Vasiliev04, Folk06, Tauber14}.
More recently, dynamical RG methods have been utilized to characterize various continuous 
phase transitions far from thermal equilibrium as well \cite{Tauber14, Kamenev11},
including the associated universal short-time or `initial-slip' relaxation features and
`aging' scaling \cite{Henkel10}.
Yet a complete classification of non-equilibrium critical points remains an open task.
  
Our study is in part motivated by recent experimental realizations of systems with strong 
light-matter coupling and a large number of degrees of freedom, which hold the potential of 
developing into laboratories for non-equilibrium statistical mechanics, and specifically for
phase transitions among distinct non-equilibrium stationary states \cite{Carusotto13}.
We mention a few but significant examples:
In ensembles of ultra-cold atoms, Bose--Einstein condensates placed in optical cavities have 
allowed experimenters to achieve strong light-matter coupling and led to the realization of 
open Dicke models \cite{Baumann10, Ritsch13}. 
The corresponding phase transition has been studied in real time, including the determination
of an associated critical exponent \cite{Brennecke13}. 
Other platforms, which hold the promise of being developed into true many-body systems by 
scaling up the number of presently available building blocks in the near future, are arrays 
of microcavities \cite{Clarke08, Hartmann08, Houck12, Koch13} and also certain optomechanical
setups \cite{Marquardt09, Chang11, Ludwig13}.
Genuine many-body ensembles in this latter class have been realized in pumped semiconductor 
quantum wells placed inside optical cavities \cite{Imamoglu96}. 
Here, non-equilibrium Bose--Einstein condensation of exciton-polaritons has been achieved 
\cite{Kasprzak06, Lagoudakis08, Roumpos12}, where the effective bosonic degrees of freedom 
result from the strong hybridization of cavity light and excitonic matter states 
\cite{Carusotto13, Moskalenko00, Keeling10}.

Two essential ingredients are shared among these non-equilibrium systems \cite{Tauber14X}. 
First, they are strongly driven by external fields and undergo a series of internal 
relaxation processes \cite{Carusotto13}. 
The irreversible non-equilibrium drive and accompanying balancing dissipation complement the 
reversible Hamiltonian dynamics, and generate both coherent and dissipative dynamics on an 
equal footing, albeit originating from physically quite independent mechanisms.
The additional irreversible terms cause manifest violations of the detailed-balance 
conditions characteristic of many-body systems in thermal equilibrium, and induce the
break-down of the equilibrium Einstein relations that connect the relaxation coefficients 
with the thermal noise strengths. 
Second, the particle number in these systems is not conserved due to the coupling of the
electromagnetic field to the matter constituents, thus opening strong loss channels for the 
effective hybridized light-matter degrees of freedom. 
The resulting quasi-particle losses must be compensated by continuous pumping in order to 
reach stable non-equilibrium stationary states.

Interestingly, detailed-balance violations generically turn out to be irrelevant for purely
relaxational critical dynamics of a non-conserved order parameter in the vicinity of 
continuous phase transitions, which are hence characterized by the equilibrium model A
universality class \cite{Tauber02}. 
Yet in systems that undergo driven-dissipative Bose-Einstein condensation, an additional 
independent critical exponent associated with the non-equilibrium drive emerges, which
describes universal decoherence at large length- and time scales; it was originally 
identified by means of a functional RG approach \cite{Sieberer13, Sieberer14}, and 
subsequently computed within the perturbative RG framework \cite{Tauber14X}.
This novel decoherence exponent should be observable in the momentum- and frequency-resolved 
single-particle response that may, e.g., be probed in homodyne detection of 
exciton-polaritons \cite{Utsunomiya08}.

In $d > 2$ dimensions, the effective dynamical description or driven-dissipative Bose
condensates utilizes a stochastic Gross--Pitaevskii equation with complex coefficients
\cite{Sieberer13, Sieberer14}, or, equivalently, a complex time-dependent Ginzburg--Landau 
equation that generalizes the equilibrium `model A' relaxational kinetics \cite{Tauber14X}.
The latter also features very prominently in the mathematical description of spontaneous
spatio-temporal pattern formation in driven non-equilibrium systems \cite{Cross93, Cross09},
and appears, for example, in the stochastic population dynamics for three cyclically 
competing species (May--Leonard model without particle number conservation) and related
spatially extended evolutionary game theory systems \cite{Frey10}.
Its critical properties have previously been investigated in the context of coupled driven
non-linear oscillators that undergo a continuous synchronization transition at a Hopf
bifurcation instability \cite{Risler05}.
Closely related mathematical models also describe stochastic wave turbulence in Langmuir
plasmas \cite{Adzhemyan89}.
Intriguingly, however, two-dimensional driven-dissipative Bose--Einstein condensation 
appears to be captured by an anisotropic variant of the Kardar--Parisi--Zhang stochastic 
partial differential equation \cite{Altman15}.

An alternative and powerful method to extract dynamical critical exponents and identify
dynamical universality classes proceeds through the analysis of non-equilibrium relaxation 
processes and the ensuing critical aging scaling \cite{Henkel10, Janssen89}.
To this end, one prepares the system initially in a fully disordered state with vanishing
order parameter, and studies its subsequent relaxation towards equilibrium or stationarity.
The initial preparation breaks time translation invariance; if the system is quenched near a
critical point, the resulting critical slowing-down renders relaxation times huge, whence
the transient non-stationary aging regime extends for very long time intervals, and this
critical initial-slip regime is governed by universal power laws \cite{Janssen89, 
Calabrese05}.
In the case of a non-conserved order parameter or equilibrium model A in the terminology of
Halperin and Hohenberg \cite{Hohenberg77}, this process is in fact characterized by an 
independent critical initial-slip exponent $\theta$ and an associated universal scaling 
function with a single non-universal scale factor \cite{Janssen89}.
We remark that the original perturbative RG treatment for the critical universal short-time
dynamics and aging scaling has only just been extended towards a non-perturbative numerical 
analysis \cite{Chiocchetta16}.

In contrast to model A relaxational kinetics, for dynamical critical systems with a conserved 
order parameter, the aging scaling regime is entirely governed by the long-time asymptotic 
stationary dynamical scaling exponents \cite{Janssen89}; this is true also for models that 
incorporate couplings to other slow conserved fields \cite{Oerding93, Janssen93}.
Consequently, the initial-slip or aging scaling regime provides a convenient means to 
quantitatively characterize critical dynamics in numerical simulations (and presumably real
experiments as well) during the system's non-equilibrium relaxation phase \cite{Zheng98}.
Non-equilibrium critical relaxation and aging scaling has also been explored in driven
systems that either display generic scale invariance, or are tuned at a continuous phase
transition point.
Prominent examples include the Kardar--Parisi--Zhang equation for driven interfaces or
growing surfaces \cite{Krech97, Daquila11, Henkel12, Odor14, Halpin14}, driven diffusive 
systems \cite{Daquila11, Daquila12}, and reaction-diffusion or population dynamics models 
that display a transition to an absorbing state, e.g., in the contact process 
\cite{Ramasco04}, and stochastic spatially extended Lotka--Volterra models for predator-prey 
competition, for which the emergence of aging scaling may serve as an early-time indicator 
for the predator species extinction \cite{Chen16}.

Inspired by these significant findings, in this present work we address the question if the
universal non-equilibrium relaxation processes in the critical complex time-dependent 
Ginzburg--Landau equation differ from the corresponding equilibrium dynamical model A~? 
In order to attack this problem mathematically, we utilize the path integral representation
of stochastic Langevin equations through a Janssen--De~Dominicis functional \cite{Janssen76, 
Dominicis76, Bausch76, Tauber14}, as previously developed and analyzed in the critical 
stationary regime for driven-dissipative Bose--Einstein condensation in 
Ref.~\cite{Tauber14X}.
Following Ref.~\cite{Janssen89} for the initial-slip and aging scaling analysis of the
relaxational models A and B in thermal equilibrium, we represent the randomized initial 
state through a Gaussian distribution for the complex-valued order parameter field.
We then employ the perturbative field-theoretical RG approach \cite{Amit84, Itzykson89, 
Kleinert01, Zinn05}, and specifically its extension to critical dynamics \cite{Vasiliev04, 
Folk06, Tauber14}, to analyze the ensuing singularities and compute the critical exponents.
Since the initial conditions at time $t = 0$ may be viewed as specifying sharp boundary
conditions on the semi-infinite time sheet, one can borrow theoretical tools originally
developed for the investigation for surface critical phenomena \cite{Diehl86}.
Near and below the upper critical dimension $d_c = 4$, the parameter $\epsilon = 4 - d$ 
serves as the effective small expansion parameter for the ensuing perturbation series in 
terms of non-linear fluctuation loops.
 
The bulk part of this paper is organized as follows: 
In the following section 2, we provide the mesoscopic dynamical model based on a stochastic 
Gross--Pitaevskii partial differential equation with complex coefficients that is motivated 
by experimental studies on driven-dissipative Bose--Einstein condensation \cite{Sieberer13,
Sieberer14}. 
Equivalently, this non-equilibrium kinetics can be viewed as relaxational model A dynamics
of a non-conserved complex order parameter field orginating from a complex-valued
Landau--Ginzburg functional \cite{Tauber14X}.
Then, utilizing the harmonic Feynman diagram components, i.e., correlation and response 
propagators that are constructed from the linear part of the associated Janssen--De~Dominicis
response functional \cite{Janssen89}, we first discuss the system's dynamics on the 
mean-field level, including the fluctuation-dissipation ratio \cite{Calabrese05}.
Section~3 details our perturbative RG calculation to lowest non-trivial (one-loop) order in
$\epsilon$. 
Upon utilizing a additional renormalization constant for the order parameter field on the
`initial-time sheet', we obtain the scaling behavior of our model and determine the 
additional independent initial-slip critical exponent associated with a fully randomized 
initial state \cite{Janssen89}. 
Through the extra renormalization constant acquired by the initial preparation that induces
breaking of time translation invariance, we extract the initial-slip exponent which governs 
the universal short-time behavior as well as the non-equilibrium relaxation in the aging 
scaling regime. 
We then proceed to discuss the resulting two-loop and higher-order corrections through 
numerical solutions of the one-loop RG flow equations for the non-linear coupling parameters 
\cite{Tauber14X} in section~4. 
In section~5, we construct a suitable complex spherical model A extension akin to 
Ref.~\cite{Henkel15} to provide an alternative demonstration for our main conclusion, namely 
that the critical aging scaling in the non-equilibrium complex Ginzburg--Landau equation is 
asymptotically governed by the equilibrium model A initial-slip exponent.   
We finally summarize our work in the concluding section~6.
A brief appendix lists the fundamental momentum loop integrals evaluated by means of the 
dimensional regularization technique that are required for the perturbative renormalization 
group calculations.

\section{Model description and mean-field analysis}

Following Refs.~\cite{Tauber14X, Sieberer13, Sieberer14}, we employ a noisy Gross--Pitaevskii
equation with complex coefficients to capture the dynamics of a Bose--Einstein condensate 
subject to dissipative losses and compensating external drive:
\begin{equation} 
    i \partial_t \psi(\bm{x},t) = \Bigl[ - (A-iD) \nabla^2 - \mu + i \chi + (\lambda-i\kappa)
	|\psi(\bm{x},t)|^2 \Bigr] \psi(\bm{x},t) + \zeta(\bm{x},t) \, . \quad
    \label{gpe}
\end{equation}
Obviously, eq.~\eref{gpe} coincides with the time-dependent complex Ginzburg--Landau 
equation, which has been prominently employed to describe pattern formation in 
non-equilibrium systems in the noise-free deterministic limit \cite{Cross93, Cross09}. 
The complex bosonic field $\psi$ here represents the polariton degrees of freedom. 
The complex coefficients have clear physical meanings as well: $\chi=(\gamma_p-\gamma_l)/2$ 
is the net gain, the balance of the incoherent pump rate $\gamma_p$ and the local 
single-particle loss rate $\gamma_l$. 
The positive parameters $\lambda$ and $\kappa$ represent the two-body loss and interaction 
strength, respectively; and $A=1/2m_{\rm eff}$ relates to the quasi-particle effective mass. 
This stochastic partial differential equation is often not presented with an explicit 
diffusion coefficient $D$, whereas a frequency-dependent pump term $\sim \eta\partial_t \psi$ 
is added on its left-hand side \cite{Wouter10L, Wouter10B}, whereupon eq.~\eref{gpe} is 
recovered through dividing by $1-i\eta$ on both sides, i.e., with $D=A\eta$ and a subleading 
correction to the other coefficients, which are complex to begin with. 
Due to the freedom of normalizing the time derivative term as above in the equation of 
motion, this model accurately captures the physics close to the phase transition, since it 
describes the most general low-frequency dynamics in a systematic derivative expansion that 
incorporates all relevant coupling in dimensions $d>2$ \cite{Tauber14X}. 
The complex Gaussian white noise term $\zeta$ can be entirely characterized through its 
correlators 
\begin{eqnarray}
    \qquad\quad\;\ \langle\zeta^*(\bm{x},t)\rangle &=& \langle\zeta(\bm{x},t)\rangle = 0 \, ,
	\nonumber\\
    \ \langle\zeta^*(\bm{x},t) \zeta(\bm{x '},t ')\rangle &=& \gamma\delta(\bm{x}-\bm{x'})
	\delta(t-t') \, , \nonumber\\
    \langle\zeta^*(\bm{x},t) \zeta^*(\bm{x '},t ')\rangle &=&\langle \zeta(\bm{x},t) 
	\zeta(\bm{x '},t ')\rangle = 0 \, .
    \label{gpnoise}
\end{eqnarray}

As mentioned above, the parameters $A$, $D$, $\lambda$, and $\kappa$ should all be positive 
for physical stability. 
On the other hand, the coefficient $\chi$ starts out negative initially and becomes positive 
as the system undergoes a continuous driven Bose--Einstein condensation transition, which
results in a non-vanishing expectation value $\langle\psi(\bm{x},t)\rangle \ne 0$. 
The parameter $\mu$, which can be considered as an effective chemical potential, needs to 
stay fixed as a requirement for stationarity. 
The Langevin equation \eref{gpe} may be obtained from a microscopic description in terms of a
quantum master equation upon employing canonical power counting in the vicinity of the 
critical point \cite{Sieberer13, Sieberer14, Gardiner99}.
For analytical convenience, we introduce the following ratios to rewrite the 
Gross--Pitaevskii equation:
\begin{equation}
    r=-\frac{\chi}{D} \, , \ r'=-\frac{\mu}{D} \, , \ u'=\frac{6\kappa}{D} \, , \
	r_K=\frac{A}{D} \, , \ r_U=\frac{\lambda}{\kappa} \ .
\end{equation}
Factoring out $iD$ on the right-hand side of \eref{gpe}, and $i\kappa$ in front of the 
non-linear term, we arrive at the equivalent stochastic partial differential equation
\begin{eqnarray}
    \partial_t\psi(\bm{x},t) &=& -D \Bigl[ r+ir'-(1+ir_K)\nabla^2+\frac{u'}{6}(1+ir_U) 
	|\psi(\bm{x},t)|^2 \Bigr] \psi(\bm{x},t) \nonumber\\
    &&+ \xi(\bm{x},t) = -D\frac{\delta {\bar H}[\psi]}{\delta \psi^*(\bm{x},t)} 
	+ \xi(\bm{x},t) \, .
\label{cgle}
\end{eqnarray}
The stochastic noise term $\xi=-i\zeta$ can be characterized similarly as $\zeta$ above. 
In the second line, we have written eq.~\eref{cgle} in the form of purely relaxational 
kinetics with a non-Hermitean effective `pseudo-Hamiltonian'
\begin{eqnarray}
    {\bar H}[\psi]&=&\int d^dx \, \Bigl[ (r+ir')|\psi(\bm{x},t)|^2
	+ (1+ir_K) |\nabla \psi(\bm{x},t)|^2 \nonumber\\
    &&\qquad\quad + \frac{u'}{12}(1+ir_U) |\psi(\bm{x},t)|^4 \Bigr] \, .
\label{cglhm}   
\end{eqnarray}

With the above assumptions, we can construct the equivalent dynamical Janssen--De~Dominicis
response functional \cite{Janssen76, Dominicis76, Bausch76} of this driven-dissipative model
by introducing a Martin--Siggia--Rose response field $\tilde{\psi}(\bm{x},t)$ to average the 
stochastic noise $\xi$ through a Gaussian integral; see, e.g., Ref.~\cite{Tauber14} for more 
detailed explanations:
\begin{eqnarray}
    A[\tilde{\psi},\psi]&=&\int\!d^dx\!\int\!dt \ \Bigg\{\tilde{\psi}^*(\bm{x},t)\Bigl[
	\partial_t + D \Big( r+ir'-(1+ir_K)\nabla^2 \Big) \Bigr] \psi(\bm{x},t) \nonumber\\
    &&\qquad\qquad\quad +\tilde{\psi}(\bm{x},t)\Bigl[ \partial_t+D \Big( r-ir'-(1-ir_K)
	\nabla^2 \Big) \Bigr] \psi^*(\bm{x},t)\nonumber\\
    &&\qquad\qquad\quad -\frac{\gamma}{2}|\tilde{\psi}(\bm{x},t)|^2+D\frac{u'}{6}(1+ir_U)
	\tilde{\psi}^*(\bm{x},t)|\psi(\bm{x},t)|^2\,\psi(\bm{x},t)\nonumber\\
    &&\qquad\qquad\quad +D\frac{u'}{6}(1-ir_U)\tilde{\psi}(\bm{x},t)|\psi(\bm{x},t)|^2\,
	\psi^*(\bm{x},t)\Bigg\} \ .
\label{jddf}
\end{eqnarray}
In addition to this bulk action \cite{Tauber14X}, we must specify randomized initial
configurations at the $t=0$ time sheet from which the system relaxes.
To this end, we assume a Gaussian weight for the initial order parameter field characterized
by
$\langle\psi(\bm{x},0)\rangle = a(\bm{x})$ at the initial time surface.
In addition to taking averages with the bulk weight $\exp({-A[\tilde{\psi},\psi]})$, we 
then require averaging with the Gaussian probability distribution 
\begin{equation}
    e^{-H_i[\psi]}=\exp \biggl[- \Delta \int\!d^dx \, |\psi(\bm{x},0)-a(\bm{x})|^2
	\biggr] \, ,
\end{equation}
which specifies an initial state with mean spatially varying order parameter $a(\bm{x})$ and
the correlations
\begin{equation}
    \big\langle\big[\psi(\bm{x},0)-a(\bm{x})\big] \big[\psi^*(\bm{x'},0)-a^*(\bm{x'})\big]
	\big\rangle = \Delta^{-1} \delta(\bm{x}-\bm{x'}) \, .
\end{equation}
We now set $\psi(\bm{x},t<0)=0$, whereupon the Gaussian part of the action \eref{jddf} 
becomes
\begin{eqnarray}
    &&A_0[\tilde{\psi},\psi] = \int\!d^dx\!\int_0^{\infty}\!dt \ \bigg\{ 
	\tilde{\psi}^*(\bm{x},t) \Bigl[ \partial_t+D \Big( r+ir'-(1+ir_K)\nabla^2 \Big) \Bigr]
	\psi(\bm{x},t) \nonumber\\
    &&\qquad\quad\ +\tilde{\psi}(\bm{x},t) \Bigl[ \partial_t+D \Big( r-ir'-(1-ir_K)\nabla^2
	\Big) \Bigr] \psi^*(\bm{x},t) -\frac{\gamma}{2}|\tilde{\psi}(\bm{x},t)|^2 \bigg\} . \
\end{eqnarray}
We finally complement the action with external source terms $J$ and $\tilde J$ conjugate to 
both the $\psi$ and ${\tilde \psi}$ fields:
\begin{eqnarray}
    A_J[\tilde{\psi},\psi] &=& -\int\!d^dx\!\int\!dt \ \Bigl[
	\tilde{J}^*(\bm{x},t)\tilde{\psi}(\bm{x},t) + J^*(\bm{x},t)\psi(\bm{x},t) \nonumber\\
    &&\qquad\qquad\quad +\tilde{J}(\bm{x},t)\tilde{\psi}^*(\bm{x},t) 
	+ J(\bm{x},t)\psi^*(\bm{x},t)\Bigr] \, .
\end{eqnarray}
Hence, the ultimate generating functional of our model becomes
\begin{equation}
    \mathcal{Z}[\tilde{J},J] = \int \mathfrak{D}[i\tilde{\psi}] \int \mathfrak{D}[\psi] \,
	\exp \Big[ -(A[\tilde{\psi},\psi]+H_i[\psi]+A_J[\tilde{\psi},\psi])\Big] \, .
\end{equation}

We first analyze the mean-field theory for our model.
By means of the Green's function technique, we may directly solve the classical field 
equations for the Gaussian generating functional $\mathcal{Z}_0[\tilde{J},J]$ to obtain the 
mean-field expressions for the expectation values $\langle\tilde{\psi}(\bm{x},t)\rangle_0$ 
and $\langle \psi(\bm{x},t)\rangle_0$:
\begin{eqnarray}
    0=\frac{\delta(A_0+H_i+A_J)}{\delta\tilde{\psi}^*(\bm{x},t)}&=&\Bigl[ 
	\partial_t+D \Big( r+ir'-(1+ir_K) \Big)\nabla^2 \Bigr] \psi(\bm{x},t)\nonumber\\
    &&-\tilde{J}(\bm{x},t)-\frac{\gamma}{2}\tilde{\psi}(\bm{x},t) \, , \nonumber\\
    0=\frac{\delta(A_0+H_i+A_J)}{\delta \psi^*(\bm{x},t)}&=&\Bigl[- \partial_t+D \Big(
	r-ir'-(1-ir_K) \Big) \nabla^2 \Bigr] \tilde{\psi}(\bm{x},t)\nonumber\\
    &&-J(\bm{x},t)-\tilde{\psi}(\bm{x},0)\delta(t)+\Delta [\psi(\bm{x},0)-a(\bm{x})]\delta(t)
	\, . \
\end{eqnarray}
The integration limit for the differential equations above is constrained to $0<t<\infty$, 
and the boundary conditions for the Martin--Siggia--Rose response field 
$\tilde{\psi}(\bm{x},t=0) = \Delta[\psi(\bm{x},0)-a(\bm{x})]$ and
$\tilde{\psi}(\bm{x},t \to \infty)=0$ are necessary to satisfy the initial distribution of 
$\psi(\bm{x},0)$. 
Solving these time differential equations in momentum space, we find for 
$\langle\tilde{\psi}(\bm{q},t)\rangle_0$ and $\langle\psi(\bm{q},t)\rangle_0$ in terms of the
conjugate sources:
\begin{eqnarray}
    \langle\tilde{\psi}(\bm{q},t)\rangle_0 &=& \int_0^\infty\!\exp \Big\{ D \Bigl[ r-ir'
	+(1-ir_K)q^2 \Big] (t-t') \Big\} \, \Theta(t-t') J(\bm{q},t') \, dt' \, , \nonumber \\
    \langle\psi(\bm{q},t)\rangle_0 &=&\int_0^\infty \exp \Big\{ -D \Big[ r+ir'+(1+ir_K)q^2
	\Bigr] (t-t') \Big\} \, \Theta(t-t') \nonumber\\
    &&\qquad\quad \times \Bigl[ \tilde{J}(\bm{q},t) + \frac{\gamma}{2}\tilde{\psi}(\bm{x},t)
	+\big[ a(\bm{q})+\Delta^{-1} \tilde{\psi}(\bm{q},t)\big] \delta(t) \Bigr] \, dt' \, .
    \label{forder}
\end{eqnarray}
Thus we determine the Gaussian response and correlation propagators
$G^0_{\tilde{\psi}^*\psi}(\bm{q},t,t')=\langle\psi^*(\bm{q},t)\tilde{\psi}(\bm{q},t')\rangle
= \delta \langle\tilde{\psi}(\bm{q},t')\rangle / \delta J(\bm{q},t)|_{J = {\tilde J} = 0}$ 
and $C^0_{\psi^*\psi}(\bm{q},t,t')=\langle\psi^*(\bm{q},t')\psi(\bm{q},t)\rangle = 
\delta \langle\psi(\bm{q},t)\rangle / \delta\tilde{J}(\bm{q},t')|_{J = {\tilde J} = 0}$,
which serve as the basic components for the perturbation expansion and Feynman diagrams.
By means of the expressions \eref{forder}, we arrive at
\begin{eqnarray}
    G^0_{\tilde{\psi}^*\psi}(\bm{q},t,t')&=&G^0_{\tilde{\psi}^*\psi}(\bm{q},t-t') 
	= e^{-D [r+ir'+(1+ir_K)q^2](t-t')} \, \Theta(t-t') \, , 
	\label{resp} \\
    C^0_{\psi^*\psi}(\bm{q},t,t')&=&C^D_{\psi^*\psi}(\bm{q},t,t') + \Delta^{-1} 
	G^0_{\tilde{\psi}^*\psi}(\bm{q},t) G^0_{\tilde{\psi}\psi^*}(\bm{q},t') \, .
    \label{corp}
\end{eqnarray}
Comparing with the bulk propagators of Ref.~\cite{Tauber14X}, the harmonic response 
propagator (\ref{resp}) here is not influenced by the initial condition and remains 
translationally invariant in time, whereas the correlation propagator (\ref{corp}), more 
precisely, its Dirichlet component $C^D_{\psi^*\psi}(\bm{q},t,t')$, distinctly reflects the 
initial preparation and does not obey time translation invariance,
\begin{equation}
    C^D_{\psi^*\psi}(\bm{q},t,t') = \frac{\gamma \, e^{-iD(r'+r_Kq^2)(t-t')}}{4D(r+q^2)} \, 
	\Bigl[ e^{-D(r+q^2)|t-t'|}- e^{-D(r+q^2)(t+t')} \Bigr] \, .
    \label{dcorp}
\end{equation}
Under RG scale transformations, the initial configuration distribution width $\Delta$ is a 
relevant parameter, and one expects $\Delta \to \infty$ under the renormalization group flow
\cite{Janssen89}.
If this asymptotic limit $\Delta \to \infty$ is taken, the second term in  
$C^0_{\psi^*\psi}(\bm{q},t,t')$ becomes eliminated, and we are left with only the Dirichlet 
correlator \eref{dcorp}. 

It is instructive to follow Ref.~\cite{Calabrese05}, and use the Gaussian response and
correlation propagators to evaluate the fluctuation-dissipation ratio
\begin{equation}
    X(\bm{q};t>t',t')= k_{\rm B}T\,
	\frac{\chi(\bm{q};t>t',t')}{dC^0_{\psi^*\psi}(\bm{q};t,t')/dt'} \, .
    \label{fdr1}
\end{equation}
In thermal equilibrium, this ratio is required to be $1$ according to Einstein's relation. 
To this end, we require the dynamic susceptibility or response function
\begin{equation}
    \chi(\bm{q};t>t',t')=D(1+ir_K)\,G^0_{\tilde{\psi}^*\psi}(\bm{q},t,t') \, ,
\end{equation}
wherefrom we obtain the inverse fluctuation-dissipation ratio \eref{fdr1} in momentum space
for our model 
\begin{eqnarray}
    &&X(\bm{q};t>t',t')^{-1} = \frac{\gamma}{4Dk_{\rm B}T(r+q^2)(1+ir_K)} \, 
	\Bigl[ r+ir'+(1+ir_K)q^2 \nonumber\\
    &&+(r-ir'+q^2-ir_Kq^2) \, e^{-2D(r+q^2)t'} \Bigr] 
	- \frac{r-ir'+(1-ir_K)q^2}{\Delta(1+ir_K)} \, e^{-2D(r+q^2)t'} \, . \
	\label{fdrc}
\end{eqnarray}
In the asymptotic time limit $t' \to \infty$, this expression reduces to
\begin{equation}
    \lim_{t' \to \infty} X(\bm{q};t>t',t')^{-1} =
	\frac{\gamma[r+ir'+(1+ir_K)q^2]}{4Dk_{\rm B}T(r+q^2)(1+ir_K)} \, .
\end{equation}
In order to satisfy the fluctuation-dissipation theorem as required for the system to relax
towards thermal equilibrium at long times, one must thus demand the following relationships 
between the parameters in the modified Gross--Pitaevskii or complex time-dependent 
Ginzburg--Landau equation \eref{cgle}:
\begin{equation}
    r' = r_K r \, , \ \gamma = 4Dk_{\rm B}T \, .
	\label{fdr2}
\end{equation}

In the critical regime $r=r'=0$ and $q^2=0$, where the characteristic relaxation time scale 
$t_c = [D(r+q^2)]^{-1}$ diverges, the fluctuation-dissipation ratio \eref{fdrc} will never 
reach the thermal equilibrium limit $1$; in fact even with equilibrium parameters \eref{fdr2}
it attains a fixed complex value at any time $t'$,
\begin{equation}
    X(0;t>t',t') = \frac{1+ir_K}{2} \, .
	\label{fdcc}
\end{equation}
In the asymptotic Dirichlet limit $\Delta \to \infty$, the fluctuation-dissipation ratio 
becomes in real space
\begin{eqnarray}
    &&X_0(\bm{x};t>t',t')^{-1} = 1+\frac{1-ir_K}{1+ir_K} 
	\Biggl( \frac{t-t'}{t+\frac{1-ir_K}{1+ir_K}\,t'} \Biggr)^{d/2} \nonumber\\
	&&\qquad\qquad\qquad \times \exp\Biggl( -2Dt' \Biggl[ r-\frac{x^2}{4D^2(t-t')(1+ir_K)^2
	\bigl(t +\frac{1-ir_K}{1+ir_K}\,t' \bigr)} \Biggr] \Biggr) \, .
\label{fdr3}
\end{eqnarray}
This result yields the corresponding equilibrium model A expression for $r_K = 0$
\cite{Calabrese05}.
In the long-time limit $t,t' \to \infty$, with the time ratio $s=t'/t$ held fixed, we find
near the critical point $r=0$,
\begin{equation}
    X_0(0;s=t'/t<1)^{-1}=1+\frac{1-ir_K}{1+ir_K} \Biggl( \frac{1-s}{1+\frac{1-ir_K}{1+ir_K}s}
	\Biggr)^{d/2} \, .
\label{fdr4}
\end{equation}
Thermal equilibrium is restored as $s\to1$. 
However, for $s=0$ the ratio \eref{fdcc} is reached: $X_0(0;0)^{-1}=1+(1-ir_K)/(1+ir_K)$. 
This suggests a crossover between the time ratio regimes $s=1$ and $s=0$, which can be 
associated with the critical initial slip exponent $\theta$. 
In the following section, we shall write down the associated general scaling laws, and
explicitly calculate $\theta$ for our specific model by means of the perturbative dynamical
RG to one-loop order, or first order in the dimensional expansion in $\epsilon=4-d$.

\section{Renormalization group analysis to one-loop order}

As established by Janssen, Schaub, and Schmittmann, the general scaling form in the 
initial-slip or critical aging regime $t' \ll t$ for the dynamical correlation function of
the equilibrium model A for a non-conserved order parameter with purely relaxational kinetics
reads
\begin{equation}
    C(\bm{q};t,t'/t \to 0)=|\bm{q}|^{-2+\eta} \, (t/t')^{\theta-1} \,
    \hat{C}_0(\bm{q}\xi,|\bm{q}|^zDt) \, ,
    \label{cor1}
\end{equation}
where $\xi \sim |\tau|^{-\nu}$ denotes the diverging correlation length as the critical
point at $\tau = 0$ is approached, with associated critical exponent $\nu$; $z$ indicates the
dynamical critical exponent that describes critical slowing-down, while $\theta$ denotes 
the universal initial-slip exponent $\theta$ \cite{Janssen89}.
In thermal equilibrium, the fluctuation-dissipation theorem then yields the corresponding
scaling form for the dynamic susceptibility:
\begin{equation}
   \chi(\bm{q};t,t'/t \to 0)=D |\bm{q}|^{z-2+\eta} \, (t/t')^\theta \,
   \hat{\chi}_0(\bm{q}\xi, |\bm{q}|^zDt) \, \Theta(t) \, .
\label{sus1}
\end{equation}
For the driven-dissipative Gross--Pitaevskii equation or complex Ginzburg--Landau equation, 
the following more general scaling form applies for the dynamical response function
\cite{Tauber14X}:
\begin{eqnarray}
   \chi(\bm{q};t,t'/t \to 0) &=& D |\bm{q}|^{z-2+\eta} \,
   (1 + i a |\bm{q}|^{\eta - \eta_c})^{-1} \, (t/t')^\theta \nonumber \\
   &&\times \hat{\chi}_0\Bigl(\bm{q}\xi, |\bm{q}|^z (1 + i a |\bm{q}|^{\eta - \eta_c})Dt 
   \Bigr) \Theta(t) \, .
\label{sus2}
\end{eqnarray}
Here, the universal correction-to-scaling exponent $\eta_c$ is induced by the external drive,
and describes the ultimate disappearance of coherent quantum fluctuations at the critical
point relative to the dissipative internal noise.
To second order in the dimensional expansion, one obtains 
$\eta_c = - [4 \ln (4/3) -1+O(\epsilon)] \eta$.
Similar additional terms apply to the dynamical correlation function \eref{cor1}, albeit in
general with also modified Fisher exponent $\eta \to \eta'$ and initial-slip exponent
$\theta \to \theta'$:
\begin{equation}
    C(\bm{q};t,t'/t \to 0)=|\bm{q}|^{-2+\eta'} \, (t/t')^{\theta'-1} \,
    \hat{C}_0(\bm{q}\xi,|\bm{q}|^zDt,a|\bm{q}|^{\eta - \eta_c}) \, .
    \label{cor2}
\end{equation}
Yet both the non-perturbative and perturbative RG analysis (already to one-loop order) have 
established that this system eventually thermalizes in the critical regime, whereupon 
detailed balance becomes effectively restored.
This thermalization, which requires that $\Delta = r_U - r_K \to 0$, implies the identities 
$\eta' = \eta$ and also $\theta' = \theta$.
In addition, asymptotically in fact $r_U = r_K \to 0$ (as established by a two-loop 
perturbative RG calculation) and hence also $r' \to 0$, whereupon eq.~\eref{cgle} turns into
the equilibrium time-dependent Ginzburg--Landau equation with a non-conserved complex order 
parameter field.
Thus the only infrared-stable RG fixed point for this stochastic dynamical system is in fact
the equilibrium model A fixed point.
Consequently the static and dynamic critical exponents $\nu$, $\eta$, and $z$ all become 
identical to those for the two-component equilibrium model A \cite{Tauber14X, Sieberer13, 
Sieberer14}.

\begin{figure}[t]
    \centering
    \includegraphics[width=5cm]{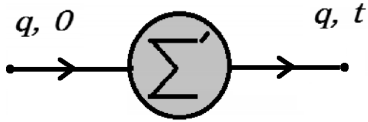}
    \caption{Full response propagator and one-particle reducible self-energy.}
    \label{fig:1}
\end{figure}
\begin{figure}[t]
    \centering
    \includegraphics[width=8cm]{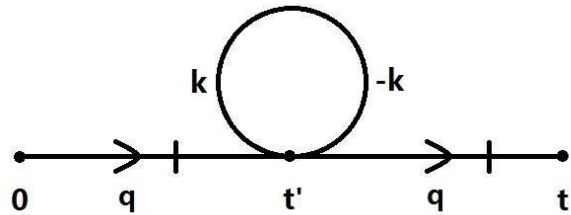}
    \caption{Feynman tadpole diagram or Hartree loop.}
    \label{fig:2}
\end{figure}

This leaves us with the explicit computation of the initial-slip or critical aging exponent 
$\theta$ for our driven-dissipative system, for which we may closely follow the procedure in
Ref.~\cite{Janssen89}. 
Hence we just sketch the essential points in this calculation.
The first step is to list the basic components for the perturbation series and associated
Feynman diagrams. 
The response \eref{resp} and correlation propagators \eref{corp} are already listed above,
and are graphically represented by directed and non-directed lines, respectively.
The non-linear fluctuation terms $\propto u'$ in the Janssen--De~Dominicis functional 
\eref{jddf} yield the four-point vertex
\begin{equation}
    -\frac12 \Gamma^0_{\tilde{\psi}\psi^*\psi^*\psi}=-D\frac{u'}{6}\,(1+ir_U)
\label{4vertex}
\end{equation}
and its complex conjugate.
The randomized initial preparation of the system breaks time translation invariance, and
induces one additional singularity that needs to be renormalized on the initial time sheet
in the temporal domain. 
Inspection of the ensuing Feynman graphs for the response propagator shows that it can
generally be written as a convolution of its stationary counterpart and a one-particle
reducible self-energy $\Sigma'$, see Fig.~\ref{fig:1}; i.e.:
\begin{equation}
    \langle\psi(-\bm{q},t)\tilde{\psi}^*(\bm{q},t)\rangle=\int_0^t \langle
	\psi(-\bm{q},t)\tilde{\psi}^*(\bm{q},t')\rangle_{\rm stat}\,\Sigma'(\bm{q},t')\,dt' \, .
\end{equation}
To first order in $u'$, the only contribution to $\Sigma'$ is the `Hartree loop' shown in
Fig.~\ref{fig:2}.
In the asymptotic limit $\Delta \to \infty$, it is to be evaluated with the Dirichlet 
correlator \eref{dcorp}, which yields
\begin{equation}
    \Sigma'(\bm{q},t)=\delta(t)-\frac23 u'D(1+ir_U)\,G_{\tilde{\psi}^*\psi}^0(\bm{q},t)
	\int\!\frac{d^dk}{(2\pi)^d}\,C_{\psi^*\psi}^D(\bm{k},t,t) \, .
    \label{selfenergy1}
\end{equation}
It is crucial to note that as the loop closes onto itself at intermediate time $t'$, the
non-equilibrium component in the first term of eq.~\eref{dcorp} that contains $r'$ and $r_K$
disappears, and the Dirichlet propagator contributions are identical to those in equilibrium.
After straightforward temporal Fourier transform, we obtain after integration with 
dimensional regularization (see appendix~A):
\begin{eqnarray}
    \Sigma'(\bm{q},\omega)&=&1+
	\frac{\gamma u'(1+ir_U)A_d}{6[(1+ir_U)r+(1+ir_K)q^2+i\omega/D](d-2)\epsilon} \nonumber\\ 
    &&\quad\ \times \frac{1}{[(3+ir_U)r/2+(1+ir_K)q^2/2+i\omega/2D]^{1-d/2}} \, , 
    \label{selfenergy2}
\end{eqnarray}
where $A_d = \Gamma(3-d/2)/2^{d-1}\pi^{d/2}$.

For the subsequent renormalization procedure, we set the normalization point to $r=0$,
$\bm{q}=0$, but $i\omega/2D=\mu^2$ outside the infrared-singular region, whence in minimal
subtraction and with \eref{fdr2} and $u = k_{\rm B}T u'$:
\begin{equation}
   \Sigma'(0,\omega)_{\rm NP}=1+\frac{u(1+ir_U)A_d\mu^{-\epsilon}}{3\epsilon} \, .
\end{equation}
Next we define the renormalization constant for the initial response field through
$\tilde{\psi_R}(\bm{x},0)=(Z_0 Z_{\tilde{\psi}})^{1/2} \tilde{\psi}(\bm{x},0)$, whence $Z_0$
absorbs the ultraviolet divergence in the renormalized self-energy:
$\Sigma_R'(\bm{q},\omega)=Z_0^{1/2}\Sigma'(\bm{q},\omega)$.
Explicitly, we then find to  one-loop order
\begin{equation}
    Z_0=1-\frac{2u_R(1+ir_{U R})}{3\epsilon}+O(u_R^2) \, ,
\end{equation}
where $u_R = Z_u u A_d\mu^{-\epsilon}$ and $r_{U R} = Z_{r_U} r_U$ with $Z_u$ and $Z_{r_U}$ 
determined in Ref.~\cite{Tauber14X}.
The associated Wilson's flow function that enters the renormalization group equation becomes
\begin{equation}
    \gamma_0(u_R)=\mu \partial_\mu|_0\ln Z_0=\frac23\,u_R(1+ir_{U R})+O(u_R^2) \, .
\end{equation}
As a final step, one resorts to a short-time expansion for the response field 
$\tilde{\psi}(\bm{x},t')=\tilde{\sigma}(t')\tilde{\psi}(\bm{x},0)+\ldots$, which through the
RG flow translates into the asymptotic scaling 
$\tilde{\sigma}(t')=(Dt')^{-\theta}\hat{\sigma}(t'/\xi^{z})$ \cite{Janssen89}, where we
identify
\begin{equation}
    \theta = \gamma_0(u^*) / 2z \, .
\end{equation}
Under the RG flow, as stated before, $r_{U R} \to 0$ \cite{Tauber14X}, and the non-linear 
coupling $u_R$ approaches an infrared-stable fixed point 
$u^* = 3 \epsilon / 5 + O(\epsilon^2)$ in dimensions $d < d_c = 4$ ($\epsilon > 0$).
Thus $\gamma_0(u^*) = 2 \epsilon / 5 + O(\epsilon^2)$, and with the standard two-loop 
critical exponents for the equilibrium model A with two order parameter components
\cite{Folk06, Tauber14}
\begin{equation}
    \eta=\epsilon^2/50+O(\epsilon^3) \, , \quad 
    z=2 + [6\ln (4/3)-1+O(\epsilon)] \, \eta \, ,
\end{equation}
we at last obtain
\begin{equation}
    \theta=\epsilon/10+O(\epsilon^2) \ ,
\end{equation}
precisely as for the two-component model A.

\section{Effect of two-loop and higher-order fluctuation corrections}

Higher-order loop corrections assuredly do not display the temporally local feature of the
tadpole graph, Fig.~\ref{fig:2}; hence non-equilibrium contributions and phase-coherent
interference terms from the dynamical correlation functions \eref{corp} cause deviations 
relative to the relaxation kinetics in the equilibrium model A.
However, we know from the one-loop RG flow equations that asymptotically all non-equilibrium
parameters flow to zero \cite{Tauber14X}.
Any effects from the coherent quantum kinetics thus ultimately disappear at the critical
point, which also applies to the critical aging scaling regime.
Yet for some initial values of the running couplings, conceivably the RG flow might 
temporarily reach a transient metastable point in parameter space, with associated dynamic
scaling properties distinct from those of the two-component model A.

\begin{figure}
    \centering
    \subfigure{\includegraphics[width=5.2cm]{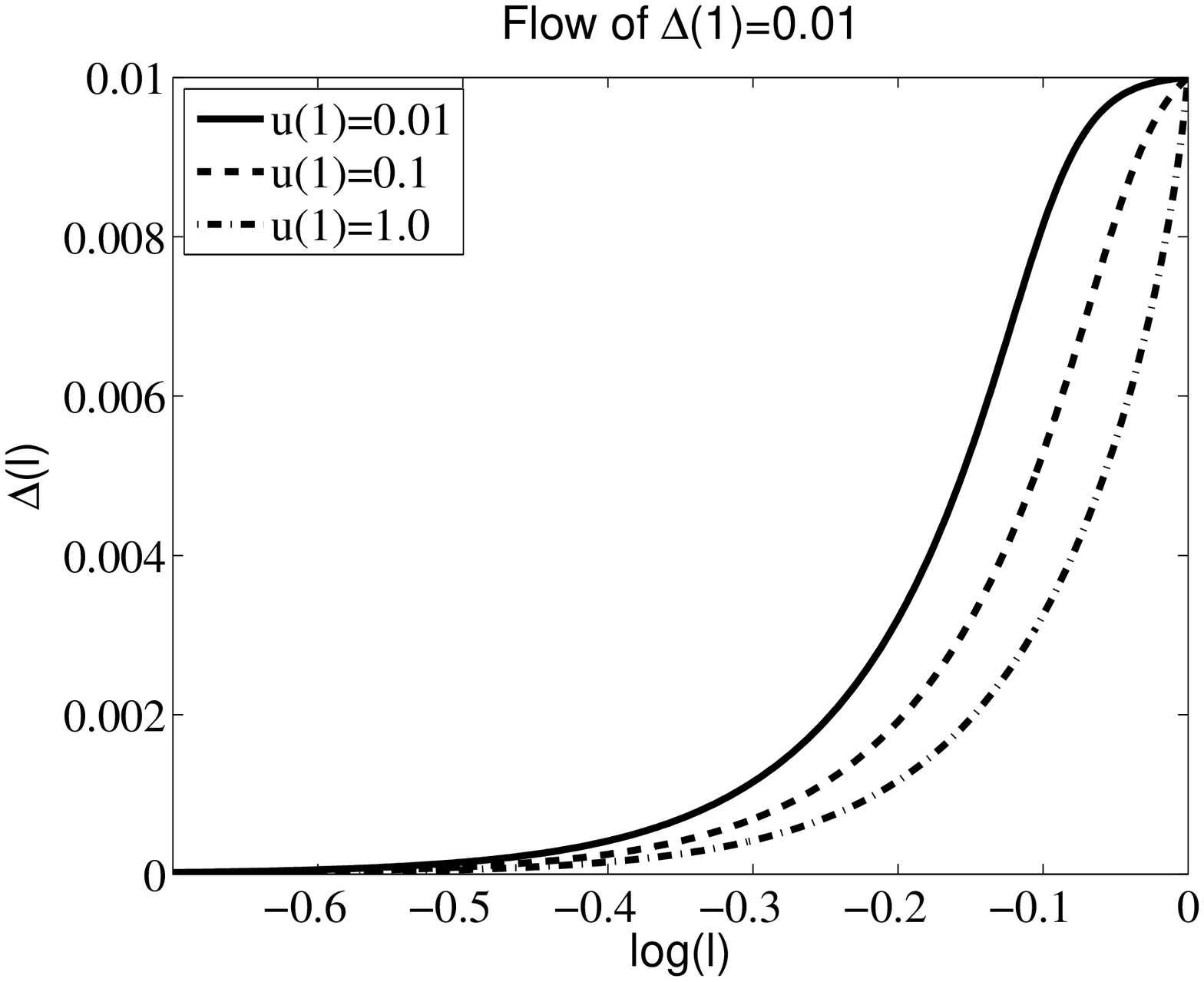}}
    \subfigure{\includegraphics[width=5.2cm]{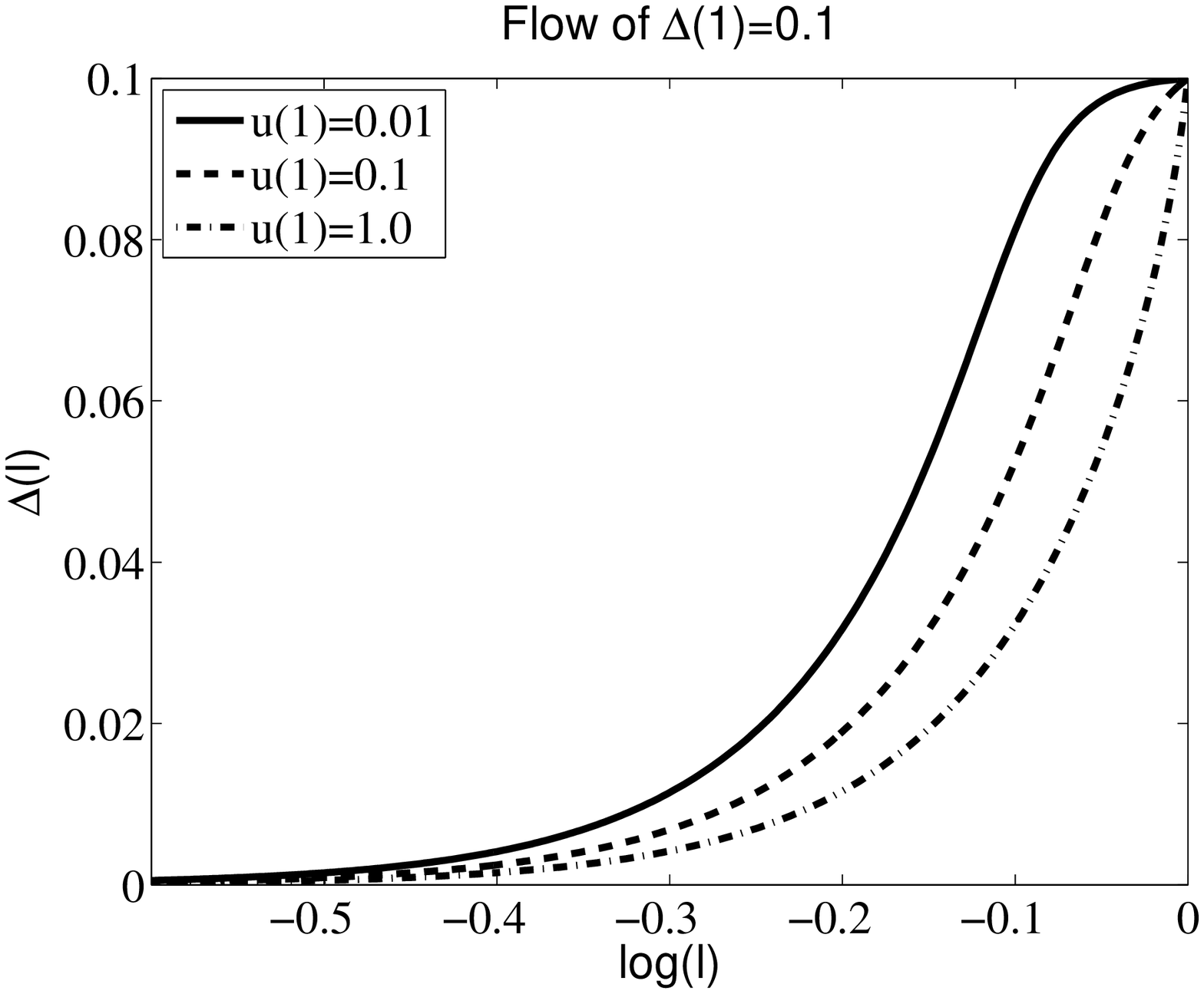}}
    \subfigure{\includegraphics[width=5.2cm]{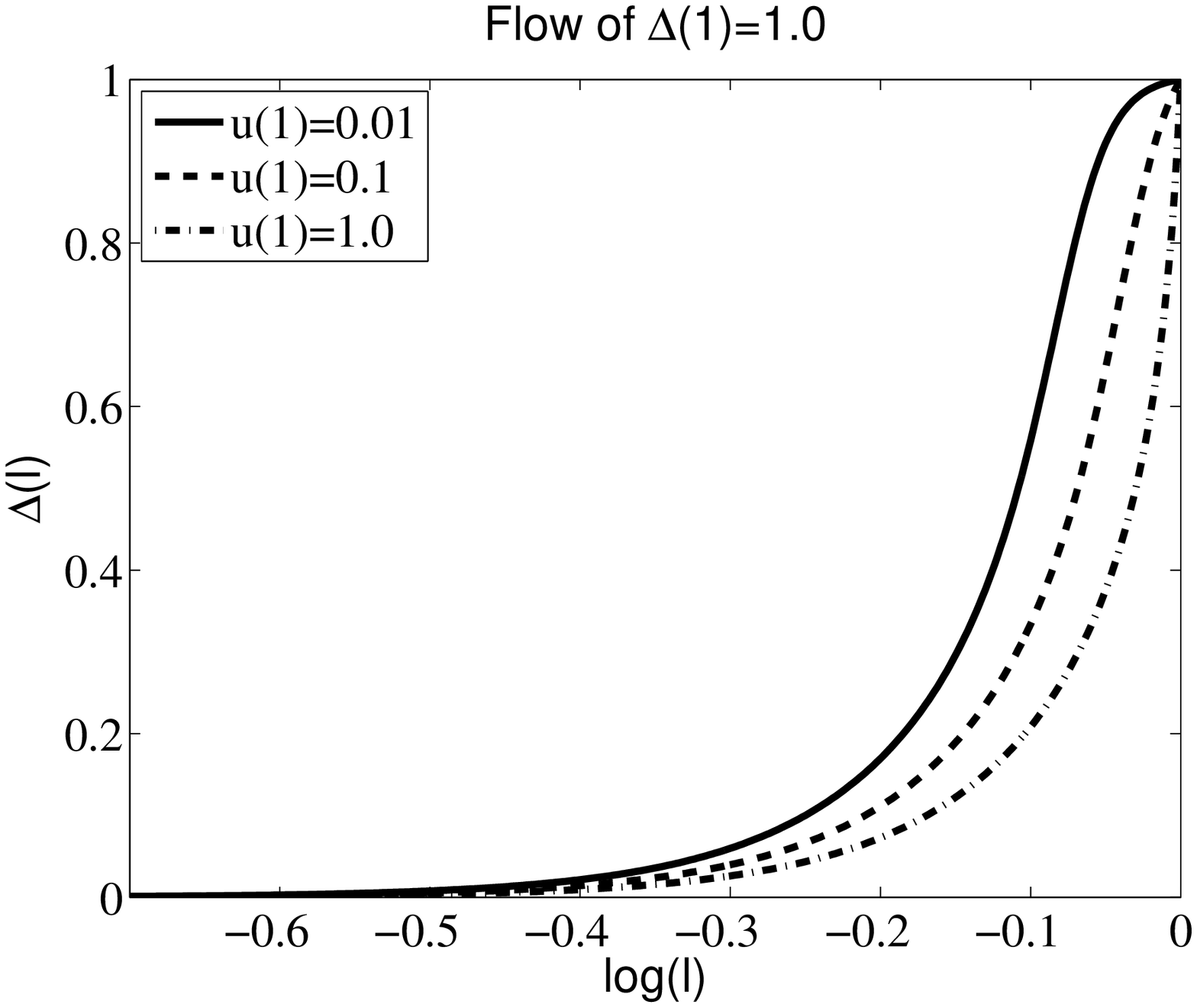}}
    \caption{Flow of the non-equilibrium parameter $\Delta(\ell)$ with initial values 
	(a) $\Delta(1)=0.01$, (b) $\Delta(1)=0.1$, and (c) $\Delta(1)=1.0$ for several different 
	initial values of the non-linear coupling $u(1)=0.01$, $0.1$, $1.0$, in $d=3$ dimensions 
	($\epsilon = 1$).}
    \label{fig:3}
\end{figure}
In order to investigate this possibility, we consider the one-loop RG flow equations for
the running counterparts of the non-linear coupling $u_R$ and the non-equilibrium parameter 
$\Delta_R = r_{U R} - r_{K R}$, as derived in Ref.~\cite{Tauber14X}:
\begin{eqnarray}
    l\partial_l u(l)&=&u(l) \Biggl[ -\epsilon + \frac53\,u(l) 
	- \frac{\Delta(l)^2}{3[1+r_K(l)^2]}\,u(l)+O\big(u(l)^2\big)\Biggr] \, , \nonumber\\
    l\partial_l \Delta(l)&=&\Delta(l) \Biggl[ 1 + \frac{2r_K(l)\Delta(l)+\Delta(l)^2}
	{1+r_K(l)^2} \Biggr] \frac{u(l)}{3}+O\big(u(l)^2\big)\, ,
\label{nlbeta}
\end{eqnarray}
obtained from the characteristics $\mu \to \mu l$.
Their ultimately stable equilibrium fixed point is $\Delta^* = 0$ and 
$u^* = 3 \epsilon / 5 + O(\epsilon^2)$.
We solve the coupled system of non-linear ordinary differential equations \eref{nlbeta} 
numerically by means of a four-step Runge-Kutta method, for various initial values $u(l=1)$
and $\Delta(l=1)$.

\begin{figure}
    \centering
    \subfigure{\includegraphics[width=5.2cm]{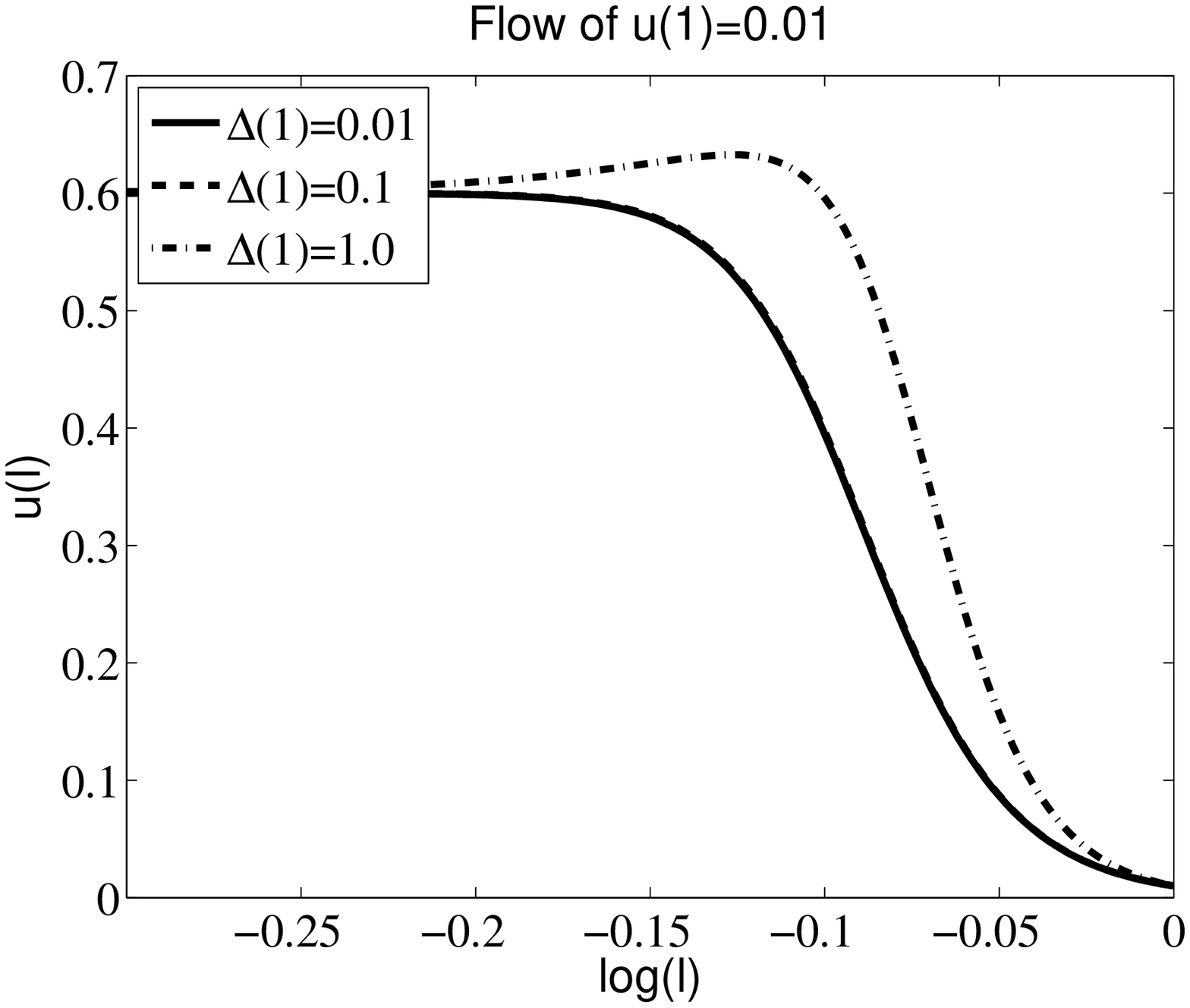}}
    \subfigure{\includegraphics[width=5.2cm]{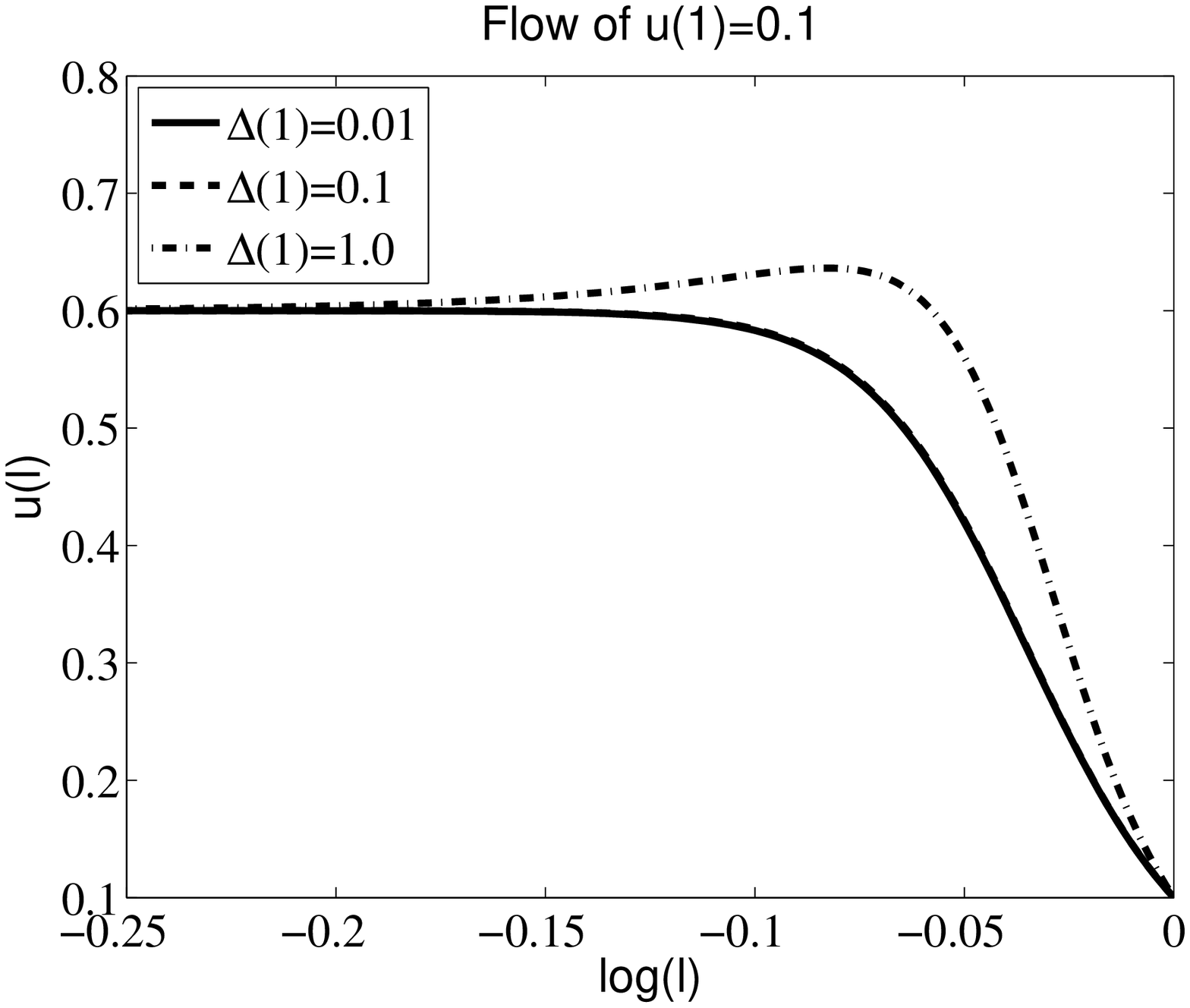}}
    \subfigure{\includegraphics[width=5.2cm]{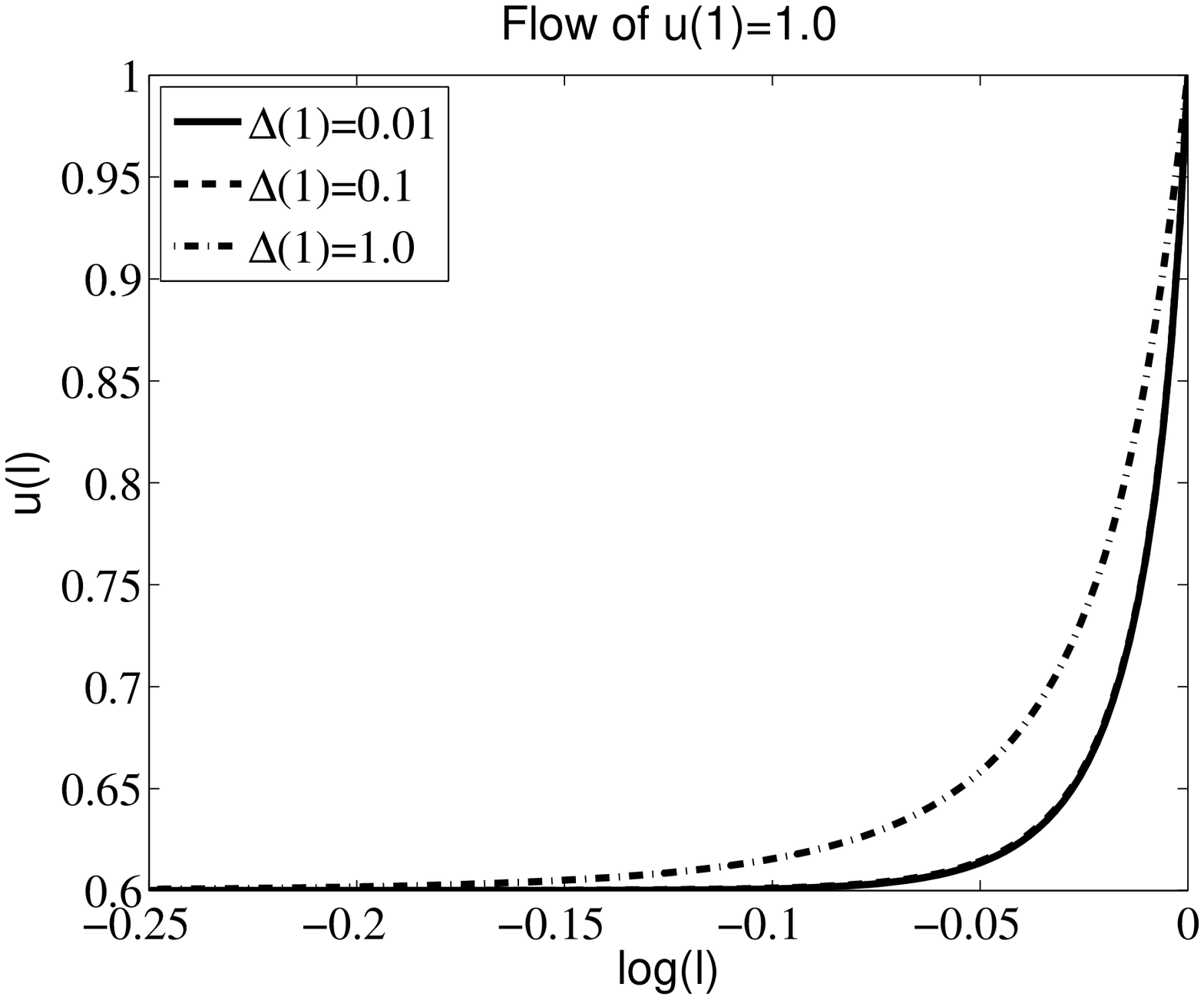}}
    \caption{Flow of the non-linear coupling parameter $u(\ell)$ with initial values 
	(a) $u(1)=0.01$, (b) $u(1)=0.1$, and (c) $u(1)=1.0$ for several different initial values 
	of the non-equilibrium parameter $\Delta(1)=0.01$, $0.1$, $1.0$, in $d=3$ dimensions 
	($\epsilon = 1$).}
    \label{fig:4}
\end{figure}
For dimensional parameter $\epsilon=1$, i.e., $d=3$, the resulting RG flows of the coupling 
parameters $\Delta(l)$ and $u(l)$ are respectively shown in Figs.~\ref{fig:3} and 
\ref{fig:4}.
We observe that the RG flow quite quickly runs into the asymptotic values $\Delta^*=0$ and
$u^*=3/5$, which represents the equilibrium model A fixed point. 
No interesting transient metastable crossover region is discernible in these graphs for
either parameter.
This leads us to anticipate that the results from two- or higher-loop fluctuation corrections
ultimately become identical to the corresponding ones for the two-component equilibrium model
A \cite{Janssen89}, and no interesting distinct crossover region emerges.

\section{Spherical model extension for the complex Ginzburg--Landau equation}

Our goal in this section is to analyze the partition function $Z[h=0]$ for an $n$-component
extension of the complex Landau--Ginzburg pseudo-Hamiltonian in the spherical model limit 
$n\to \infty$, which can be directly generated from \eref{cglhm}:
\begin{eqnarray}
    {\bar H}[\psi^\alpha]&=&\!\int\!d^dx\,\Bigl[ (r+ir')\sum_{\alpha=1}^n 
	|\psi^\alpha(\bm{x})|^2 + (1+ir_K) \sum_{\alpha=1}^n|\nabla \psi^\alpha(\bm{x})|^2 
	\nonumber\\
    &&\qquad\quad + \frac{u'}{12}(1+ir_U) \Bigl( \sum_\alpha|\psi^\alpha(\bm{x})|^2 \Bigr)^2 
	\, \Bigr] \, .
\label{ncglhm}   
\end{eqnarray}
The corresponding spherical equilibrium model A has been investigated extensively in 
previous work, utilizing either a self-consistent decoupling method \cite{Janssen89} or a 
Gaussian Hubbard--Stratonovich transformation to effectively `linearize' the quartic 
non-linear term in this Hamiltonian (see, e.g., Refs.~\cite{Tauber14, Henkel10, Henkel15}. 
These two approaches are equivalent, but we employ the latter to analyze our non-equilibrium 
system. 
To this end, we introduce an auxiliary field $\Psi(\bm{x})$, through which the Gaussian 
Hubbard--Stratonovich transformation can be performed,
\begin{equation}
    \int d(i\Psi) \, e^{-(1+ir_U)[\Psi(\sum_\alpha|\psi^\alpha|^2)-3\Psi^2/u']}\propto 
	e^{-(1+ir_U)(\sum_\alpha|\psi^\alpha|^2)^2/12} \, .
    \label{ghstr}
\end{equation}
Substituting this transformation as well as $r'=r_K r$, the augmented Hamiltonian becomes
\begin{eqnarray}
    {\tilde H}[\psi^\alpha,\Psi]&=&\!\int\!d^dx\,\Bigl[ (1+ir_K)[r+\Psi(\bm{x})]
	\sum_\alpha|\psi^\alpha(\bm{x})|^2
	+ (1+ir_K) \sum_\alpha|\nabla \psi^\alpha(\bm{x})|^2 \nonumber\\
    &&\qquad\quad - \frac{3}{u'}(1+ir_U)\Psi(\bm{x})^2 \Bigr] \, .
\label{ncglh1}   
\end{eqnarray}
At this point the original order parameter fields can be integrated out, and one arrives at
\begin{eqnarray}
    Z[h=0]\!&\propto&\!\!\int\!\!\mathcal{D}[i\Psi] \exp\Bigl[\frac{3(1+ir_U)}{u'}
	\Psi(\bm{x})^2 -n \Tr \ln \frac{(1+ir_K)G_\Psi(\bm{x},\bm{x'})^{-1}}{2\pi}\Bigr] ,
\end{eqnarray}
with the inverse Green's function
\begin{equation}
    G_\Psi(\bm{x},\bm{x'})^{-1}=[r+\Psi(\bm{x})-\nabla^2]\,\delta(\bm{x}-\bm{x'}) 
\end{equation}
and its Fourier transform in momentum space
\begin{equation}
    G_\Psi(\bm{q},\bm{q'})^{-1}=(r+q^2)\,(2\pi)^d\delta(\bm{q}+\bm{q'})+\Psi(\bm{q}+\bm{q'})
	\, .
\end{equation}

Now recall that the RG fixed point for the non-linear coupling is 
$u^* \propto \epsilon/(n+2)$; thus, for large $n$, resetting the non-linear coupling as
$u=u'/n$ will render the parameter $u'$ independent of the number of components $n$ as 
$n\to\infty$. 
This yields the partition function $Z[h=0]\propto\int\mathcal{D}[i\Psi]e^{-n\Phi[\Psi]}$ with
the effective potential
\begin{equation}
    \Phi[\Psi]=-\frac{3(1+ir_U)}{u'}\int\!d^dx \, \Psi^2(\bm{x})
	+ \Tr \ln \frac{(1+ir_K)G_\Psi(\bm{x},\bm{x'})^{-1}}{2\pi} \, .
\end{equation}
In the spherical model limit $n\to\infty$, the steepest-descent approximation will become 
exact, that is we need to seek the solution of the classical field equation 
$\delta\Phi[\Psi]/\Psi(\bm{x})=0$.
For simplicity, we assume a homogeneous solution $\Psi(\bm{x})=\Psi$, whence the stationarity
condition yields a self-consistent equation for $\Psi$:
\begin{equation}
    \Psi=\frac{u'}{6(1+ir_U)}\int\!\frac{d^dq}{(2\pi)^d}\,\frac{1}{r+\Psi+q^2} \, .
\end{equation}
This result looks precisely like its equibilirium spherical model A counterpart, aside from 
the overall complex prefactor $1 + i r_U$.
Specifically, the integral is just the bare correlation function $C_0(\bm{x}=0)$ with a 
shifted temperature parameter $r\to r+\Psi$. 
Yet previous work \cite{Tauber14} established that asymptotically $r_U\to 0$ at the stable
RG fixed point; therefore, one obtains the static critical exponents of the equilibrium 
spherical model A: $\eta=0$ and $\gamma=2\nu=2/(d-2)$ for $d < d_c=4$. 
Furthermore, the analysis for the dynamics of this non-equilibrium system will also be 
essentially identical as for the equilibrium spherical model \cite{Tauber14, Henkel10}, and
results in the dynamical critical and initial-slip exponents
\begin{equation}
    z = 2 \, , \ \theta=(4-d)/4 \ ,
\end{equation}
both coinciding identical with the equilibrium spherical model A values.

The above analysis of the spherical model extension for the time-dependent complex 
Ginzburg--Landau equation of course holds to all orders in a perturbative expansion. 
The fact that the spherical model limit too recovers the equilibrium values for all critical
exponents of this system further supports our conclusion in the previous section~4 that 
higher-order fluctuation corrections to the critical initial-slip exponent for our driven
non-equilibrium kinetics must be identical to those of model A in thermal equilibrium.

\section{Conclusion and outlook}

We have investigated the driven-dissipative non-equilibrium critical dynamics of a 
non-conserved complex order parameter field.
Specifically, we have addressed the situation where the system experiences a sudden change in 
its parameters that quenches it from a random initial configuration into the critical regime. 
We have mainly focused on the initial-slip critical exponent $\theta$ which governs the 
universal short-time behavior during the transient non-equilibrium relaxation period before 
the asymptotic long-time stationary regime is reached. 
We have employed the perturbative field-theoretical renormalization group method to calculate
the value of $\theta$ to first order in the dimensional $\epsilon$ expansion. 
Our explicit result turns out identical to that for the equilibrium dynamical model A 
\cite{Janssen89}. 
Quantum coherence effects do not modify this universal scaling exponent owing to the temporal
locality of the one-loop Feynman diagram, or equivalently the fact that the phase term in the
correlation propagator is annihilated rendering the results identical to those for the 
equilibrium system without drive.
Rather than analytically calculating the complicated higher-order loop corrections, we have
invoked the one-loop renormalization group flow equations \cite{Tauber14X} as well as a 
suitable spherical model extension, constructed along the lines of Ref.~\cite{Henkel15}, to 
argue that the above conclusion likely remains true to all orders in the perturbation 
expansion.

In the future, we intend to study this and related stochastic dynamical systems by means of 
direct numerical integration.
Comparing the resulting data with our analytical theory should further aid our quantitative 
understanding of the dynamical critical properties of driven-dissipative quantum systems that 
experience parameter quenches, and hence take us another step closer towards the ultimate 
goal of obtaining a complete and systematic classification of non-equilibrium dynamical 
criticality.

\ack 
The authors are indebted to Sebastian Diehl, Andrea Gambassi, Hannes Janssen, Michel 
Pleimling, and Lukas Sieberer for helpful discussions, and to Hiba Assi for a careful 
critical reading of the manuscript draft.
This research is supported by the U.S. Department of Energy, Office of Basic Energy Sciences, 
Division of Materials Science and Engineering under Award DE-FG02-09ER46613.

\appendix

\section{Dimensional regularization, Feynman parametrization}

In order to arrive at a small expansion parameter for the perturbational analysis in our
field-theoretic RG approach, we need to consider non-integer spatial dimensions close and 
below the upper critical dimension $d_c = 4$.
We may consider these non-integer dimensionalities as an analytical continuation of integer 
ones by means of dimensional regularization. 
The fluctuation loop integrals in momentum space associated with the Feynman diagrams are
tpyically of the following form (see, e.g., Ref.~\cite{Tauber14}):
\begin{equation}
    I^{(\sigma,s)}_d(\tau) = \int \frac{d^d k}{(2\pi)^d} \frac{k^{2\sigma}}{(\tau+k^2)^s} 
	= \frac{\Gamma(\sigma+d/2) \Gamma(s-\sigma-d/2)}
	{2^d\pi^{d/2} \Gamma(d/2) \Gamma(s)} \, \tau^{\sigma-s+d/2} \ .
\end{equation}
Integrals with different denominators can be reduced to this form through Feynman's 
parametrization:
\begin{equation}
    \frac{1}{A^r B^s} = \frac{\Gamma(r+s)}{\Gamma(r)\Gamma(s)}
	\int^1_0 \frac{x^{r-1}(1-x)^{s-1}}{[xA+(1-x)B]^{r+s}} \, dx \ .
\end{equation}
These expressions are widely used to evaluate the momentum loop integrals in Sec.~2.
Euler's gamma function provides the appropriate interpolation for non-integer dimensions.


\section*{References}
\begin{thebibliography}{99}

\bibitem{Amit84}
  Amit D J 1984
  {\it Field Theory, the Renomalization Group, and Critical Phenomenona}
  (Singapore: World Scientific)
   
\bibitem{Itzykson89}
  Itzykson C and Drouffe J-M 1989
  {\it Statistical Field Theory:} Vols.~I, II 
  (Cambridge: Cambridge University Press)

\bibitem{Kleinert01}
  Kleinert H and Schulte-Frohlinde V 2001
  {\it Critical Properties of $\phi^4$ Theories}
  (Singapore: World Scientific)

\bibitem{Zinn05}
  Zinn-Justin J 2005
  {\it Quantum Field Theory and Critical Phenomenona}
  (Singapore: World Scientific) 4th ed.
   
\bibitem{Hohenberg77}
  Hohenberg P C and Halperin B I 1977
  {\it Rev. Mod. Phys.} {\bf 49} 435

\bibitem{Vasiliev04}
  Vasil'ev A N 2004
  {\it The Field Theoretical Renormalization Group om Critical Behavior Theory and 
  Stochastic Dynamics} (Boca Raton: Chapman \& Hall / CRC)

\bibitem{Folk06} 
  Folk R and Moser G 2006
  {\it J. Phys. A: Math. Gen.} {\bf 39} R207  

\bibitem{Tauber14}
  T\"auber U C 2014
  {\it Critical Dynamics -- A Field Theory Approach to Equilibrium and
  Non-Equilibrium Scaling Behavior}
  (Cambridge: Cambridge University Press)

\bibitem{Kamenev11}
  Kamenev A 2011
  {\it Field Theory of Non-equilibrium Systems}
  (Cambridge: Cambridge University Press)

\bibitem{Henkel10}
  Henkel M and Pleimling M 2010
  {\it Non-equilibrium Phase Transitions, Vol. 2: Ageing and Dynamical Scaling Far from 
  Equilibrium} (Dordrecht: Springer) 

\bibitem{Carusotto13}
  Carusotto I and Ciuti C 2013
  {\it Rev. Mod. Phys.} {\bf 85} 299
  
\bibitem{Baumann10}
  Baumann K, Guerlin C, Brennecke F and Esslinger T 2010
  {\it Nature (London)} {\bf 464} 1310
  
\bibitem{Ritsch13}
  Ritsch H, Domokos P, Brennecke F and Esslinger T 2013
  {\it Rev. Mod. Phys.} {\bf 85} 553

\bibitem{Brennecke13}
  Brennecke F, Mottl F R, Baumann K, Landig R, Donner T and Esslinger T 2013
  {\it Proc. Natl. Acad. Sci. U.S.A.} {\bf 110} 11 763
  
\bibitem{Clarke08}
  Clarke J and Wilhelm F K 2008
  {\it Nature (London)} {\bf 453} 1031
  
\bibitem{Hartmann08}
  Hartmann M J, Brand\"ao F G S L and Plenio M B 2008
  {\it Laser Photonics Rev.} {\bf 2} 527
  
\bibitem{Houck12}
  Houck A A, T\"ureci H E and Koch J 2012
  {\it Nat. Phys.} {\bf 8} 292

\bibitem{Koch13}
  Schmidt S and Koch J 2013
  {\it Ann. Phys. (Berlin)} {\bf 525} 395

\bibitem{Marquardt09}
  Marquardt F and Girvin S M 2009
  {\it Physics} {\bf 2} 40

\bibitem{Chang11}
  Chang D E, Safavi-Naeini A H, Hafezi M and Painter O 2011
  {\it New J. Phys.} {\bf 13} 023003
 
\bibitem{Ludwig13}
  Ludwig M and Marquardt F 2013
  {\it Phys. Rev. Lett.} {\bf 111} 073603
 
\bibitem{Imamoglu96}
  Imamoglu A, Ram R J, Pau S and Yamamoto Y 1996
  {\it Phys. Rev. A} {\bf 53} 4250
 
\bibitem{Kasprzak06}
  Kasprzak J, Richard M, Kundermann S, Baas A, Jeambrun P, Keeling J M J, Marchetti F M, 
  Szymanska M H, Andre R, Staehli J L, Savona V, Littlewood P B, Deveaud B and Dang le S 2006
  {\it Nature (London)} {\bf 443} 409
  
\bibitem{Lagoudakis08}
  Lagoudakis K G, Wouters M, Richard M, Baas A, Carusotto I, Andre R, Dang L S and 
  Deveaud-Pl\'edran B 2008
  {\it Nat. Phys.} {\bf 4} 706
   
\bibitem{Roumpos12}
  Roumpos G, Lohse M, Nitsche W H, Keeling J, Szymanska M H, Littlewood P B, L\"offler A, 
  H\"ofling' S, Worschech L, Forchel A and Yamamoto Y 2012
  {\it Proc. Natl. Acad. Sci. U.S.A.} {\it 109} 6467
   
\bibitem{Moskalenko00}
  Moskalenko S A and Snoke D W 2000
  {\it Bose--Einstein Condensation of Excitons and Biexcitons}
  (Cambridge: Cambridge University Press)
 
\bibitem{Keeling10}
  Keeling J, Szymanska M H and Littlewood P B 2010 
  in {\it Optical Generation and Control of Quantum Coherence in Semiconductor 
  Nanostructures}, eds. Slavcheva G and Roussignol P (Berlin: Springer)
 
\bibitem{Tauber14X}
  T\"auber U C 2014 and Diehl S 2014
  {\it Phys. Rev. X} {\bf 4} 021010

\bibitem{Tauber02}
  T\"auber U C, Akkineni V K and Santos J E 2002 
  {\it Phys. Rev. Lett.} {\bf 88} 045702
  
\bibitem{Sieberer13}
  Sieberer L M, Huber S D, Altman E and Diehl S 2013
  {\it Phys. Rev. Lett.} {\bf 110} 195301
 
\bibitem{Sieberer14}
  Sieberer L M, Huber S D, Altman E and Diehl S 2014
  {\it Phys. Rev. B} {\bf 89} 134310
  
\bibitem{Utsunomiya08}
  Utsunomiya S, Tian L, Roumpos G, Lai C W, Kumada N, Fujisawa T, Kuwata-Gonokami M, 
  L\"offler A, H\"ofling S, Forchel A and Yamamoto Y 2008
  {\it Nat. Phys.} {\bf 4} 700

\bibitem{Cross93}
  Cross M C and Hohenberg P C 1993
  {\it Rev. Mod. Phys.} {\bf 65} 851

\bibitem{Cross09}
  Cross M and Greenside H 2009
  {\it Pattern Formation Outside of Equilibrium Systems}
  (Cambridge: Cambridge University Press)
  
\bibitem{Frey10}
  Frey E 2010
  {\it Physica A} {\bf 389} 4265
 
\bibitem{Risler05}
  Risler T, Prost J and J\"ulicher F 2005
  {\it Phys. Rev. E} {\bf 72} 016130

\bibitem{Adzhemyan89}
  Adzhemyan L Ts, Vasil'ev A N, Gnatich M and Pis'mak Yu M 1989
  {\it Theor. Math. Phys.} {\bf 78} 260
  
\bibitem{Altman15}
  Altman E, Sieberer L M, Chen L, Diehl S and Toner J 2015
  {\it Phys. Rev. X} {\bf 5} 011017 
  
\bibitem{Janssen89}
  Janssen H K, Schaub B and Schmittmann B 1989
  {\it Z. Phys. B} {\bf 73} 539

\bibitem{Calabrese05} 
  Calabrese P and Gambassi A 2005
  {\it J. Phys. A: Math. Gen.} {\bf 38} R133
  
\bibitem{Chiocchetta16}
  Chiocchetta A, Gambassi A, Diehl S, and Marino J 2016
  {\it e-print} {\tt arXiv:1606.06272}
  
\bibitem{Oerding93} 
  Oerding K and Janssen H K 1993
  {\it J. Phys. A: Math. Gen.} {\bf 26} 3369

\bibitem{Janssen93} 
  Oerding K and Janssen H K 1993
  {\it J. Phys. A: Math. Gen.} {\bf 26} 5295  
  
\bibitem{Zheng98}
  Zheng B 1998
  {\it Int. J. of Mod. Phys. B} {\bf 12} 1419  
  
\bibitem{Krech97}
  Krech M 1997 
  {\it Phys. Rev. E} {\bf 55} 668  

\bibitem{Daquila11}
  Daquila G L and T\"auber U C 2011
  {\it Phys. Rev. E} {\bf 83} 051107  
  
\bibitem{Henkel12}
  Henkel M, Noh J D and Pleimling M 2012
  {\it Phys. Rev. E} {\bf 85} 030102(R)  

\bibitem{Odor14}
  \'Odor G, Kelling J and Gemming S 2014
  {\it Phys. Rev. E} {\bf 89} 032146
  
\bibitem{Halpin14}
  Halpin-Healy T and Palasantzas G 2014  
  {\it EPL} {\bf 105} 50001
  
\bibitem{Daquila12} 
  Daquila G L and T\"auber U C 2012
  {\it Phys. Rev. Lett.} {\bf 108} 110602

\bibitem{Ramasco04}
  Ramasco J J, Henkel M, Santos M A and da~Silva~Santos C A 2004 
  {\it J. Phys. A Math. Gen.} {\bf 37} 10497 

\bibitem{Chen16}
  Chen S and T\"auber U C 2016
  {\it Phys. Biol.} {\bf 13} 025005
  
\bibitem{Janssen76}
  Janssen H K 1976
  {\it Z. Phys. B} {\bf 23} 377
  
\bibitem{Dominicis76}
  De~Dominicis C 1976
  {\it J. Phys. (Paris)} Colloq. {\bf 37} C1-247  

\bibitem{Bausch76}
  Bausch R, Janssen H K and Wagner H 1976
  {\it Z. Phys. B} {\bf 24} 113

\bibitem{Diehl86}
  Diehl H W 1986
  in {\it Phase Transitions and Critical Phenomena}, Vol. 10, eds. Domb C and Lebowitz J L 
  (London: Academic Press), p. 75  
  
\bibitem{Henkel15}
  Henkel M and Durang X
  {\it J. Stat. Mech.} P05022 (2015)   

\bibitem{Wouter10L}
  Wouters M and Carusotto I 2010
  {\it Phys. Rev. Lett.} {\bf 105} 020602
  
\bibitem{Wouter10B}
  Wouters M, Liew T C H and Savona V 2010
  {\it Phys. Rev. B} {\bf 82} 245315 

\bibitem{Gardiner99}
  Gardiner C W and Zoller P 1999
  {\it Quantum Noise}
  (Berlin: Springer) 

\end{thebibliography}
\end{document}